\documentclass[aps,pra,twocolumn]{revtex4}
\usepackage{graphicx}
\usepackage{dcolumn}
\usepackage{bm}
\usepackage{amsmath}
\usepackage{amssymb,amsthm}
\usepackage{epsfig,color}
\usepackage[sans]{dsfont}

\definecolor{nblue}{rgb}{0.2,0.2,0.7}
\definecolor{ngreen}{rgb}{0.2,0.6,0.2}
\definecolor{nred}{rgb}{0.7,0.2,0.2}
\definecolor{nblack}{rgb}{0,0,0}

\newcommand{\tr}{\text{tr}}

\def\tr{\mbox{tr}}

\def\bea{\begin{eqnarray}}
\def\eea{\end{eqnarray}}
\begin{document}
\title{Geometric discord of quantum states of fermionic system in accelerated frame}
\author{Jinho Chang$^{1}$}
\author{L.C. Kwek$^{2}$}
\author{Younghun Kwon$^{1}$}

\email{yyhkwon@hanyang.ac.kr}

\affiliation{${ }^{1}$Department of Physics, Hanyang University,
Ansan, Kyunggi-Do, 425-791, South Korea\\
${ }^{2}$ Center of Quantum Technology, National University of
Singapole, 3 Science Drive 2, Singapole 117543 }

\date{\today}

\begin{abstract}
In this article, we investigate the geometric discord of quantum
states of fermionic system in accelerated frame. It is shown by the
method beyond the single-mode approximation, depending on the region considered, that the geometric
discord for the entangled quantum states of fermionic system in
accelerated frame can vanish or be retained at the infinite acceleration
limit: it does not disappear when the quantum state of the particle(Alice)-particle(Bob in
region I) case or the particle(Alice)-antiparticle(Bob in region II)
is considered and it disappears when the
particle(Alice)-antiparticle(Bob in region I) case or the
particle(Alice)-particle(Bob in region II) one is considered.
\end{abstract}

\maketitle

\section*{I.Introduction}

 Quantum information has allowed us to elucidate quantum physics on the basis of
 information theory. The understanding of correlation in quantum state has been a central
theme in the study.  Insights regarding entangled quantum
state naturally lead us to apply the basic concepts in quantum
information to quantum communication. A well known
example is quantum cryptography.  A basic resource for
quantum communication is an entangled quantum state, which
cannot be obtained by local operations and classical
communications(LOCC)
between parties who share the quantum states.\\

 From another perspective, quantum information in non-inertial frame has also
 been studied \cite{ref:alsing1}\cite{ref:terashima1}\cite{
ref:terashima2}\cite{ref:ball}\cite{ref:fuentes}\cite{ref:versteeg}\cite{ref:alsing2}\cite{ref:martin}.
When the measurement based quantum state in an accelerated frame is
considered, one of most important aspects about the quantum state is
that the entangled quantum states of bosonic system in an
accelerated frame have vanishing entanglement at the infinite
acceleration limit however the entangled quantum states of fermionic
system in an accelerated frame have non-zero entanglement in the
same
limit.\\

The quantum discord is regarded as another way to understand the
correlation of quantum states. Discord is obtained through the
quantum correlation after removing classical correlation so that the
resulting correlation displays only the pure correlation with
quantumness. Studying quantum discord of a quantum state in an
accelerated frame can be important for a better understanding
quantum correlation of a state\cite{ref:datta}\cite{ref:brown}.
Specifically, the behavior of quantum states at the infinite
acceleration limit should be understood since such analysis could
provide insight into how quantum states behave near black holes for
instance. Discord of the quantum state of fermionic system in an
accelerated frame has been studied within single-mode approximation
though the latter may not always hold for all physical systems. In
this article, we investigate the behavior of discord for quantum
states of fermionic system in accelerated frame beyond single-mode
approximation\cite{ref:montero2}\cite{ref:montero3}\cite{ref:chang}.

There are actually different ways to compute quantum discord. In
our study, we restrict our measure to  geometric discord. It is found that the
geometric discord for entangled quantum states of fermionic system
in accelerated frame need not vanish even at the inifinite
acceleration limit when the particle(Alice)-particle(Bob in region
I) case and the particle(Alice)-antiparticle(Bob in region II) one
are considered. However, the geometric discord for entangled quantum
states of fermionic system in accelerated frame disappears at the
inifinite acceleration limit for
particle(Alice)-antiparticle(Bob in region I)
and particle(Alice)-particle(Bob in region II) cases.\\

The rest of this article is described as follows: In section II, we provide a brief
description to physics of an accelerated frame. In section
III, we investigate quantum discord for quantum states of fermionic system in accelerated
frame. Finally, in section IV, we discuss and summarize our
results.

\section*{II Accelerated Frame}

An accelerated frame is best described by invoking Rindler coordinates
 $(\tau,\varsigma,y,z)$, instead of employing the usual Minkowski coordinates $(t,x,y,z)$.
 Rindler coordinates can be written as
\begin{eqnarray}
ct&=&\varsigma \sinh(\frac{a \tau}{c}),x=\varsigma \cosh(\frac{a
\tau}{c}) \\
ct&=&-\varsigma \sinh(\frac{a \tau}{c}),x=-\varsigma \cosh(\frac{a
\tau}{c})
\end{eqnarray}
where $a$ denotes the fixed acceleration of the frame and $c$ is the
velocity of light. In fact, Eq(1) covers only the right wedge(called
region I) and Eq(2) describes the left wedge(called region II).\\
 The field in Minkowski and Rindler spacetime is written as
\begin{eqnarray}
&\phi & = N_{M}\sum_{i}(a_{i,M}v^{+}_{i,M} +
b^{\dag}_{i,M}v^{-}_{i,M} )\nonumber\\
& = & N_{R}\sum_{j}(a_{j,I}v^{+}_{j,I} + b^{\dag}_{j,I}v^{-}_{j,I} +
a_{j,II}v^{+}_{j,II} + b^{\dag}_{j,II}v^{-}_{j,II} )
\end{eqnarray}
Here $a^{\dag}_{i,\Delta}(a_{i,\Delta})$ and
$b^{\dag}_{i,\Delta}(b_{i,\Delta})$, which satisfy the
anticommutation relations, are the creation(annihilation) operators
for the positive and negative energy solutions(particle and
antiparticle) and $\Delta $ denotes $M,I,II$. A combination of
Minkowski mode, called Unruh mode, can be transformed into single
Rindler mode and can annihilate the same Minkowski vacuum,
satisfying the relation
\begin{equation}
A_{\Omega,R/L}\equiv \cos r_{\Omega}a_{\Omega,I/II} - \sin r_{\Omega}
b^{\dag}_{\Omega,II/I}
\end{equation}
where $\cos r_{\Omega}=(e^{\frac{-2 \pi \Omega c}{a}}+1)^{-1/2}$.
However we may obtain more general relation such as
\begin{equation}
a^{\dag}_{\Omega,U}=q_{L}(A^{\dag}_{\Omega ,L}\otimes I_{R}) +
q_{R}(I_{L} \otimes  A^{\dag}_{\Omega ,R}),
\end{equation}
beyond the single mode approximation. Using this
relation, in the case of Grassmann scalar, the Unruh vacuum and the
one-particle state is given by
\begin{eqnarray}
|0_{\Omega }\rangle_{U} &=& \cos^{2} r_{\Omega } |0000\rangle_{\Omega
} - \sin r_{\Omega } \cos r_{\Omega } |0110\rangle_{\Omega }
\nonumber\\
                        &+& \sin r_{\Omega } \cos r_{\Omega } |1001\rangle_{\Omega } - \sin^{2}
r_{\Omega } |1111\rangle_{\Omega } \nonumber\\
|1_{\Omega }\rangle^{+}_{U} &=& q_{R}(\cos r_{\Omega }
|1000\rangle_{\Omega } - \sin r_{\Omega } |1110\rangle_{\Omega })
\nonumber\\
                        &+& q_{L}(\sin r_{\Omega } |1011\rangle_{\Omega } + \cos
 r_{\Omega } |0010\rangle_{\Omega }
\end{eqnarray}
where we have used the notation $|pqmn\rangle_{\Omega } \equiv |p_{\Omega
}\rangle^{+}_{I}|q_{\Omega }\rangle^{+}_{II}  |m_{\Omega
}\rangle^{-}_{II} |n_{\Omega }\rangle^{-}_{I} $. Here we consider
$q_{R}$ and $q_{L}$ as real number. Throughout this paper,  we study within the fermionic structure
which also underlies physical framework proposed by
 \cite{ref:montero2,ref:montero3,ref:chang}.

\section*{III Geometric discord of quantum state}

 The quantum discord which was first introduced by Zurek {\it et al} for isolating quantum
correlation from classical correlation\cite{ref:henderson}
\cite{ref:ollivier}. The quantum correlation is obtained by
considering the difference between two different ways of writing
quantum mutual information. Quantum mutual information of quantum
state shared by Alice and Bob can be given by
\begin{equation}
I(A:B)=S(\rho_{A})+S(\rho_{B})-S(\rho_{AB})
\end{equation}
where $S(\rho)=-\tr(\rho \log_{2}\rho) $ denotes the von Neumann
entropy.  Quantum mutual information can also be described by
\begin{equation}
J(A:B)=\mbox{max}_{\{\Pi_{i}\}}(S(\rho_{B})-S_{\{\Pi_{i}\}}(B|A))
\end{equation}
where  $\{\Pi_{i}\}$ means a complete set of projection operator.
Quantum discord is defined as
\begin{equation}
D(A:B)=I(A:B)-J(A:B),
\end{equation}
which can be interpreted as a pure quantum correlation of quantum state.
Quantum discord can be also defined in a geometric way in which one can regard
discord as the minimum distance between the state to the set of zero discord
 states $\chi$\cite{ref:dakic}:
\begin{equation}
D_{G}=\mbox{\rm min}_{\chi \in \Omega_{0} }\| \rho_{AB}-\chi \|^{2} \nonumber
\end{equation} where
$\Omega_{0}$ denotes the set of zero discord states and $\|\| $
is the squared Hilbert-Schmidt norm.  For computational convenience, the geometric discord
is shown to be equivalent to
\begin{equation}
D_{G}=\frac{1}{4}(\|\overrightarrow{x}\|^{2}+\|T\|^{2} -
\mbox{\rm max} [\mbox{ \rm eigenvalues} (\overrightarrow{x}\overrightarrow{x}^{t}+TT^{t})]
\end{equation} where $x_{i}=\tr(\rho \sigma_{i} \otimes I)$ and $T_{ij}=\tr(\rho
\sigma_{i}\otimes\sigma_{j})$ and $t$ denotes the transpose of vectors or matrices.\\

\subsection*{A.  Two-party entangled state}
We consider a generalized $\Phi^{+}$ state such as
\begin{equation}
|\Phi^{+}\rangle=\cos \alpha |00\rangle + \sin \alpha|11^{+}\rangle \label{eq10}
\end{equation}
Suppose two parties, Alice and Bob, initially prepare a generalized $\Phi^{+}$ state in
an inertial frames and thereafter Bob moves in an accelerated frame.
Since Bob has unaccessible part due to his acceleration, the physical state that Alice
and Bob(using particle in Bob's region I) may share is given by
$\rho^{\Phi^{+}}_{AB_{I}^{+}}$, if we go
beyond the single mode approximation.
Likewise, we denote the
physical state of Alice's particle and Bob's antiparticle in Bob's
region I, Alice's particle and Bob's particle in Bob's region II,
and Alice's particle and Bob's antiparticle in Bob's region II as
$\rho^{\Phi^{+}}_{AB_{I}^{-}}$, $\rho^{\Phi^{+}}_{AB_{II}^{+}}$, and
$\rho^{\Phi^{+}}_{AB_{II}^{-}}$ respectively. The geometric discord
of $ \rho_{AB_{\mathrm{I}}^{+}}^{\Phi_{+}}$, $
\rho_{AB_{\mathrm{I}}^{-}}^{\Phi_{+}}$, $
\rho_{AB_{\mathrm{II}}^{+}}^{\Phi_{+}}$ and $
\rho_{AB_{\mathrm{II}}^{-}}^{\Phi_{+}}$ can be shown to be
\begin{widetext}
\begin{eqnarray}
D_{G}(\rho_{AB_{\mathrm{I}}^{+}}^{\Phi_{+}}) &=& \frac{1}{4}(\cos ^2
2\alpha +(\cos 2 \gamma \cos ^2 \alpha + (q_{R}^{2}-q_{L}^{2} \cos
2\gamma) \sin^{2} \alpha )^{2} +q_{R}^{2} \cos ^{2}\gamma \sin ^2
2\alpha)+  q_{R}^{2} \cos ^{2}\gamma \sin ^2
2\alpha \nonumber\\
&-& \mbox{max} [\cos ^2 2\alpha + (\cos ^2 \alpha \cos 2\gamma
+(q_{R}^{2}-q_{L}^{2}\cos 2\gamma) \sin^{2} \alpha )^{2}, q_{R}^{2}
\cos ^{2}\gamma \sin ^2 2\alpha] \nonumber\\
D_{G}(\rho_{AB_{\mathrm{I}}^{-}}^{\Phi_{+}}) &=& \frac{1}{4}(\cos ^2
2\alpha + (\cos ^2 \alpha \cos 2\gamma -(q_{L}^{2}+q_{R}^{2}\cos
2\gamma) \sin^{2} \alpha)^{2} + 2 q_{L}^{2} \cos ^{2}\gamma \sin^{2}
\gamma \sin ^2 2\alpha \nonumber\\
&-& \mbox{max} [\cos ^2 2\alpha + (\cos ^2 \alpha \cos 2\gamma
-(q_{L}^{2}+q_{R}^{2}\cos 2\gamma) \sin^{2} \alpha )^{2}, q_{L}^{2}
\cos ^{2}\gamma \sin^{2} \gamma \sin ^2 2\alpha] \nonumber\\
D_{G}(\rho_{AB_{\mathrm{II}}^{+}}^{\Phi_{+}}) &=& \frac{1}{4}(\cos ^2
2\alpha  + (\cos ^2 \alpha \cos 2\gamma -(q_{L}^{2}-q_{R}^{2}\cos
2\gamma) \sin^{2} \alpha )^{2} + 2
q_{L}^{2} \cos ^{2}\gamma \sin^{2} \gamma \sin ^2 2\alpha \nonumber\\
&-& \mbox{max} [\cos ^2 2\alpha + (\cos ^2 \alpha \cos 2\gamma
-(q_{L}^{2}-q_{R}^{2}\cos 2\gamma) \sin^{2} \alpha )^{2}, q_{L}^{2}
\cos ^{2}\gamma sin^{2} \gamma \sin ^2 2\alpha] \nonumber\\
D_{G}(\rho_{AB_{\mathrm{II}}^{-}}^{\Phi_{+}}) &=& \frac{1}{4}(\cos ^2
2\alpha  +  q_{R}^{2} \cos ^{2}\gamma \sin ^2 2\alpha  + (\cos 2 \gamma
\cos ^2 \alpha - (q_{R}^{2}+q_{L}^{2}\cos 2\gamma) \sin^{2} \alpha
)^{2} + 2 q_{R}^{2} \sin ^{2}\gamma \sin ^2 2\alpha)\nonumber\\
&-& \mbox{max} [\cos ^2 2\alpha + (\cos ^2 \alpha \cos 2\gamma -
(q_{R}^{2}+q_{L}^{2}\cos 2\gamma) \sin^{2} \alpha )^{2}, q_{R}^{2} \sin
^{2}\gamma \sin ^2 2\alpha]
\end{eqnarray}
\end{widetext}
The geometric discord for the quantum states in the various regimes
can be numerically evaluated and the behavior with acceleration is
shown in Fig 1- 2. Fig. 1 illustrates the quantum discord for the
quantum states $\rho_{AB_{\mathrm{I}}^{+}}^{\Phi_{+}}$  and $
\rho_{AB_{\mathrm{II}}^{-}}^{\Phi_{+}}$ and Fig. 2 shows the quantum
discord for the quantum states $
\rho_{AB_{\mathrm{I}}^{-}}^{\Phi_{+}}$ and $
\rho_{AB_{\mathrm{II}}^{+}}^{\Phi_{+}}$.  From the plots, we see
that the quantum discord for $\rho_{AB_{\mathrm{I}}^{+}}^{\Phi_{+}}$
and $ \rho_{AB_{\mathrm{II}}^{-}}^{\Phi_{+}}$ never vanish even at
the infinite acceleration, which implies that quantum states $
\rho_{AB_{\mathrm{I}}^{+}}^{\Phi_{+}}$ and $
\rho_{AB_{\mathrm{II}}^{-}}^{\Phi_{+}}$ have non-zero quantum
correlation even at the infinite acceleration. In case of $
\rho_{AB_{\mathrm{II}}^{+}}^{\Phi_{+}}$ and $
\rho_{AB_{\mathrm{I}}^{-}}^{\Phi_{+}}$ their quantum discord
disappear at the infinite acceleration, meaning that there is no
quantum correlation of $ \rho_{AB_{\mathrm{II}}^{+}}^{\Phi_{+}}$ and
$ \rho_{AB_{\mathrm{I}}^{-}}^{\Phi_{+}}$ in the limit of  infinite
acceleration. It should be noted that as we have seen in the
entanglement behavior of $ \rho_{AB_{\mathrm{I}}^{+}}^{\Phi_{+}}$, $
\rho_{AB_{\mathrm{I}}^{-}}^{\Phi_{+}}$, $
\rho_{AB_{\mathrm{II}}^{+}}^{\Phi_{+}}$, and $
\rho_{AB_{\mathrm{II}}^{-}}^{\Phi_{+}}$, the quantum discord of
quantum state $ \rho_{AB_{\mathrm{I}}^{+}}^{\Phi_{+}}$($
\rho_{AB_{\mathrm{I}}^{-}}^{\Phi_{+}}$) coincides with that of $
\rho_{AB_{\mathrm{II}}^{-}}^{\Phi_{+}}$($
\rho_{AB_{\mathrm{II}}^{+}}^{\Phi_{+}}$) at the infinite
acceleration.

\begin{figure}[h!]
\begin{center}
\includegraphics[width=7cm]{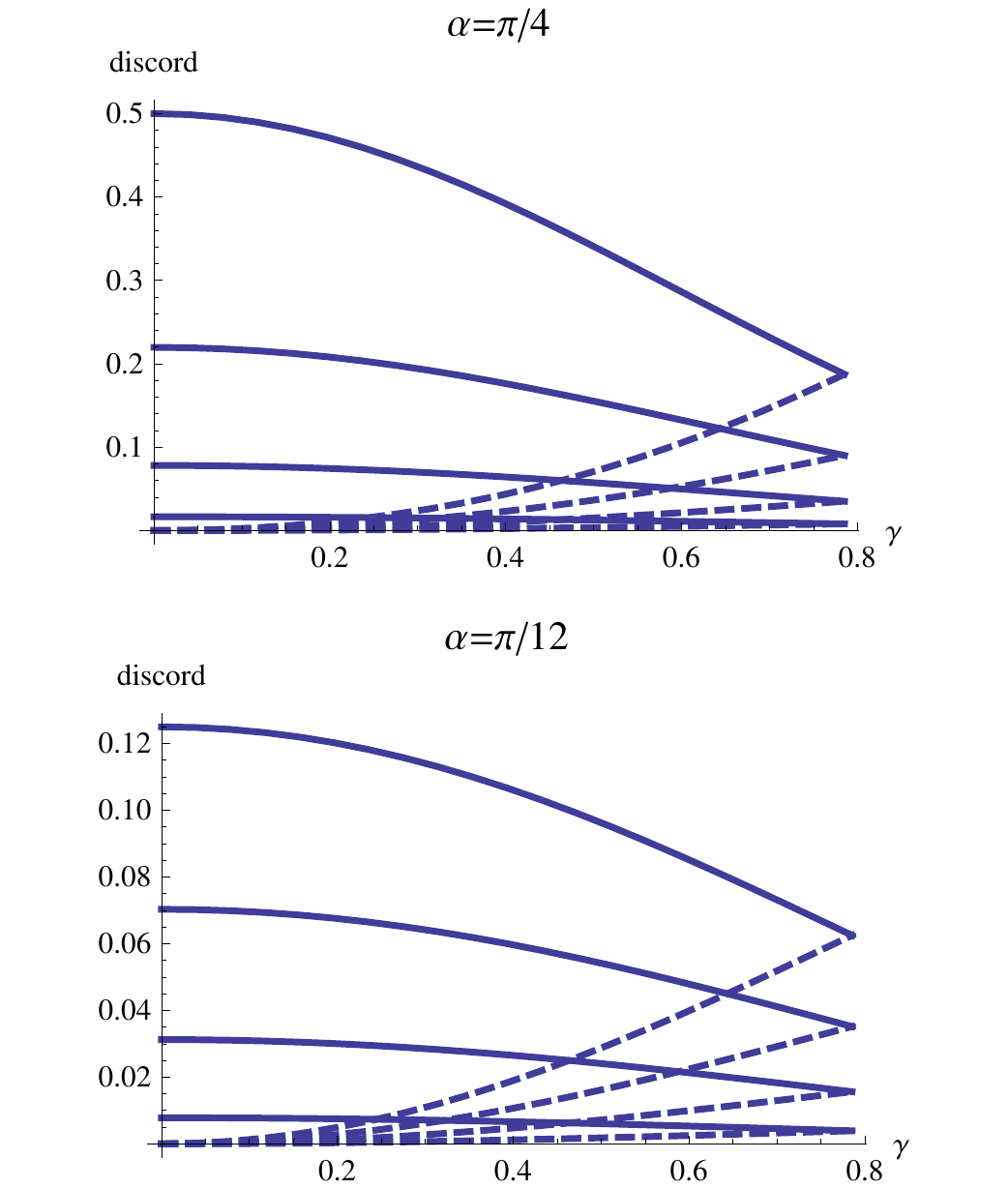}
\caption{\label{fig2}(Color online) The geometric discord of quantum
states $ \rho_{AB_{\mathrm{I}}^{+}}^{\Phi_{+}}$ and $
\rho_{AB_{\mathrm{II}}^{-}}^{\Phi_{+}}$ when $\alpha=\frac{\pi}{4}$
and $\alpha=\frac{\pi}{12}$. The solid(dotted) lines (from top to
bottom) denote the geometric discord of $
\rho_{AB_{\mathrm{I}}^{+}}^{\Phi_{+}}$($
\rho_{AB_{\mathrm{II}}^{-}}^{\Phi_{+}}$) at $q_{R}=1$, $q_{R}=0.75$
$q_{R}=0.5$ and $q_{R}=0.25$ respectively. The quantum states such
as  $ \rho_{AB_{\mathrm{I}}^{+}}^{\Phi_{+}}$ and $
\rho_{AB_{\mathrm{II}}^{-}}^{\Phi_{+}}$ show the non-vanishing
geometric discord even at the infinite acceleration. Here
$\gamma=\frac{\pi}{4}$ denotes the infinite acceleration. }
\end{center}
\end{figure}

\begin{figure}[h!]
\begin{center}
\includegraphics[width=7cm]{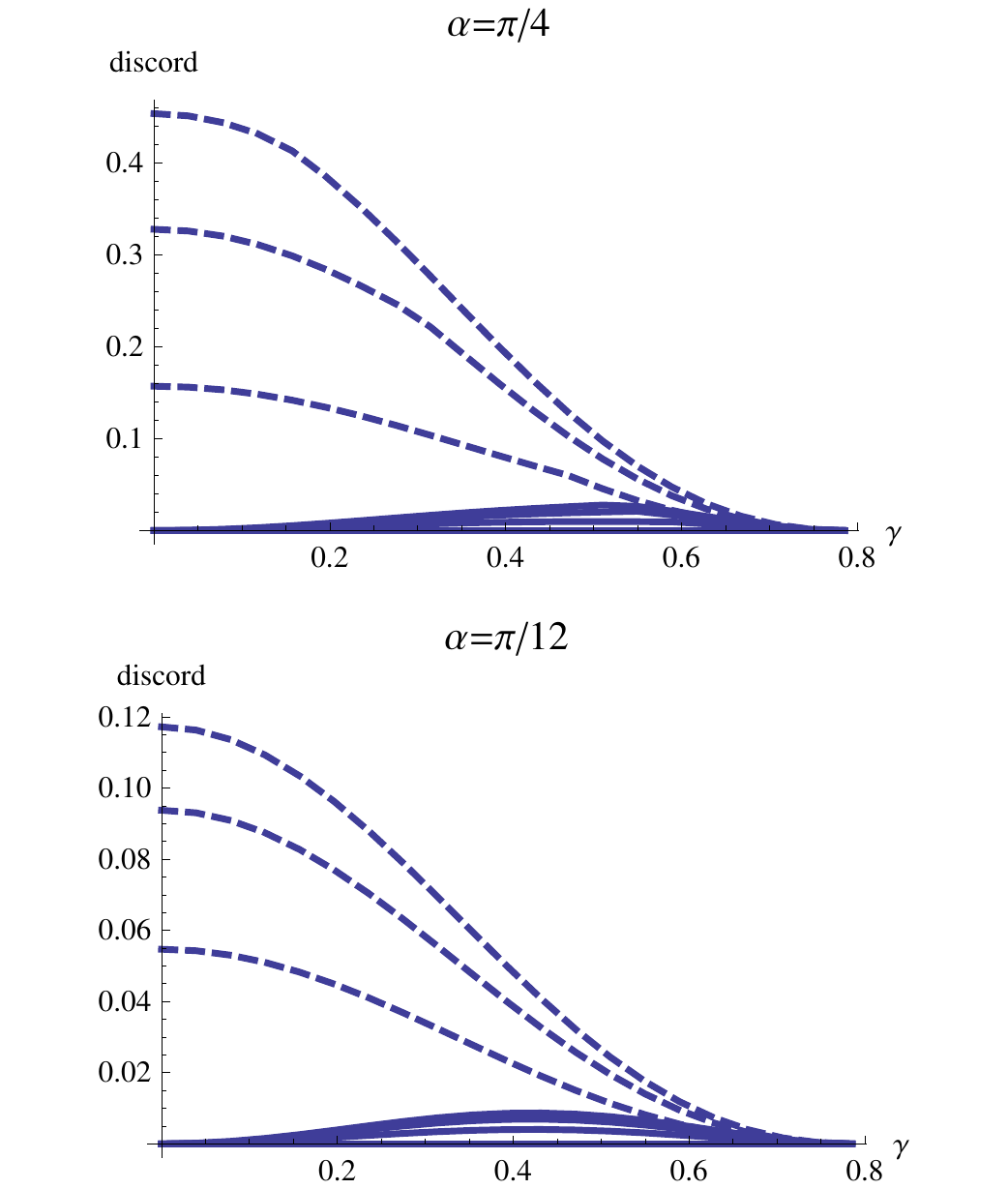}
\caption{\label{fig2}(Color online) The geometric discord of quantum
states $\rho_{AB_{\mathrm{I}}^{-}}^{\Phi_{+}}$ and $
\rho_{AB_{\mathrm{II}}^{+}}^{\Phi_{+}}$ when $\alpha=\frac{\pi}{4}$
and $\alpha=\frac{\pi}{12}$. The solid(dotted) lines (from top to
bottom) denote the geometric discord of
$\rho_{AB_{\mathrm{I}}^{-}}^{\Phi_{+}}$ and $
\rho_{AB_{\mathrm{II}}^{+}}^{\Phi_{+}}$ at $q_{R}=1$, $q_{R}=0.75$
$q_{R}=0.5$ and $q_{R}=0.25$ respectively. However the quantum
discords of the quantum states such as  $
\rho_{AB_{\mathrm{I}}^{-}}^{\Phi_{+}}$ and $
\rho_{AB_{\mathrm{II}}^{+}}^{\Phi_{+}}$ disappear at the infinite
acceleration. Here $\gamma=\frac{\pi}{4}$ denotes the infinite
acceleration. }
\end{center}
\end{figure}

So far we have considered the one-particle state described by $|1_{\Omega }\rangle^{+}_{U}$.
The one-particle state may also take the form
\begin{eqnarray}
|1_{\Omega }\rangle^{-}_{U} &=& q_{L}(\cos \gamma_{\Omega }
|0100\rangle_{\Omega } - \sin \gamma_{\Omega } |0111\rangle_{\Omega
})
\nonumber\\
                        &+& q_{R}(\sin \gamma_{\Omega } |1110\rangle_{\Omega } +\cos
 \gamma_{\Omega } |0010\rangle_{\Omega }
\end{eqnarray}
In this case, one has a generalized $\Phi^{-}$ state given by
\begin{equation}
|\Phi^{-}\rangle=\cos \alpha |00\rangle +\sin \alpha|11^{-}\rangle
\end{equation}
As in the previous analysis, suppose that two parties Alice and Bob prepare the new
generalized
$\Phi^{-}$ state in inertial frames and Bob moves in an
accelerated frame afterwards.
Since Bob has unaccessible part due to his
acceleration, if we go beyond the single mode approximation,the
physical state $\rho^{\Phi^{-}}_{AB_{I}^{+}}$ that Alice and
Bob(using particle in Bob's region I) share is obtained by
tracing out all other parts. In this way,  one obtains the
physical state $\rho^{\Phi^{-}}_{AB_{I}^{-}}$ of Alice and Bob(using
anti-particle in Bob's region I), the physical state
$\rho^{\Phi^{-}}_{AB_{II}^{+}}$ of Alice and antiBob(using particle
in Bob's region II), and the physical state
$\rho^{\Phi^{-}}_{AB_{II}^{-}}$ of Alice and antiBob(using
anti-particle in Bob's region II). Using
$\rho^{\Phi^{-}}_{AB_{I}^{+}}$,$\rho^{\Phi^{-}}_{AB_{I}^{-}}$,$\rho^{\Phi^{-}}_{AB_{II}^{+}}$
and $\rho^{\Phi^{-}}_{AB_{II}^{-}}$, one can compute the mutual
information, classical communication and geometric discord.
 It turns out that the behavior of quantum discord
 for $\rho_{AB_{I}^{+}}^{\Phi^{-}}$,$\rho_{AB_{I}^{+}}^{\Phi^{-}}$,$\rho_{AB_{I}^{+}}^{\Phi^{-}}$
 and $\rho_{AB_{I}^{+}}^{\Phi^{-}}$ is the same
 as that of $\rho_{AB_{I}^{+}}^{\Phi^{+}}$,$\rho_{AB_{I}^{+}}^{\Phi^{+}}$,$\rho_{AB_{I}^{+}}^{\Phi^{+}}$
 and $\rho_{AB_{I}^{+}}^{\Phi^{+}}$ respectively.

\subsection*{B.  Two-party mixed entangled state}
 Until now, we have considered quantum discord of pure states in
fermionic system when one of parties travels with a uniform
acceleration. We next briefly consider a more general scenario when the two
parties share a mixed state.  Naturally, we find that how the behavior of
quantum discord depends on the degree of mixedness. More specifically,  we consider
a Werner state, where a white noise component is added to a maximally
entangled states. Suppose two parties Alice and Bob initially prepare
a Werner state \begin{equation}
\rho_{W}= F |\Phi^{+} (\alpha=\pi/4) \rangle  \langle \Phi^{+}
(\alpha=\pi/4) | + \frac{1-F}{4}\mathbb{I}\label{eq9},
\end{equation}
in inertial frame where the maximally entangled state $|\Phi^{+} (\alpha=\pi/4) \rangle$
corresponds to $\alpha=\pi/4$ in
Eq. (\ref{eq10}), and then Bob moves in an
uniformly accelerated frame.

\begin{figure}[h!]
\begin{center}
\includegraphics[width=7cm]{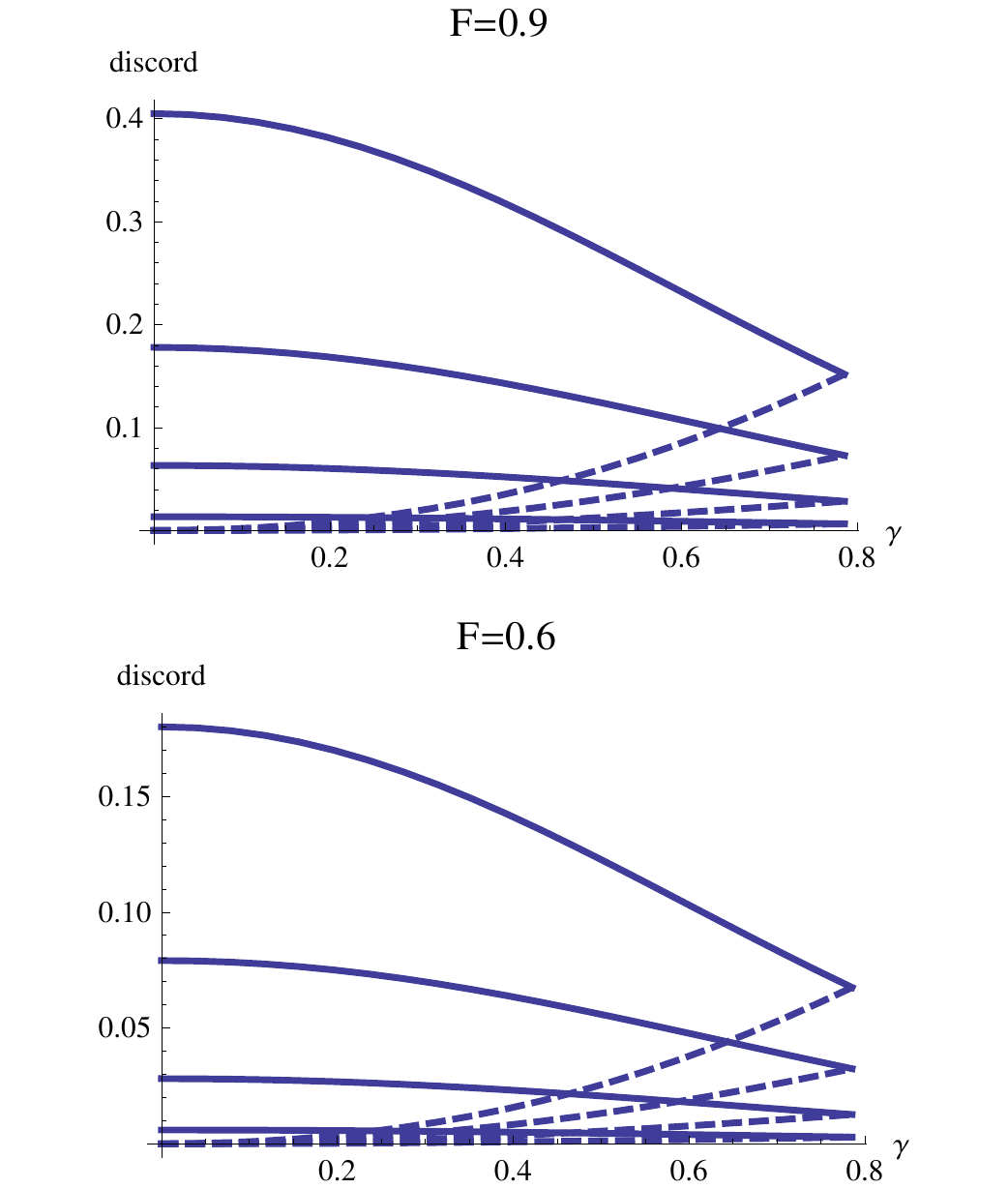}
\caption{\label{fig2}(Color online) The geometric discord of quantum
states $\rho_{AB_{\mathrm{I}}^{+}}^{W}$ and $
\rho_{AB_{\mathrm{II}}^{-}}^{W}$, when $F=0.9$ and $F=0.6$. The
solid(dotted) lines (from top to bottom) denote the geometric
discord of $ \rho_{AB_{\mathrm{I}}^{+}}^{W}$($
\rho_{AB_{\mathrm{II}}^{-}}^{W}$) at $q_{R}=1$, $q_{R}=0.75$
$q_{R}=0.5$ and $q_{R}=0.25$ respectively. The quantum states such
as $\rho_{AB_{\mathrm{I}}^{+}}^{W}$ and $
\rho_{AB_{\mathrm{II}}^{-}}^{W}$ show the non-vanishing geometric
discord even at the infinite acceleration. Here
$\gamma=\frac{\pi}{4}$ denotes the infinite acceleration. }
\end{center}
\end{figure}

Since Bob has unaccessible part due to his acceleration,
the physical state shared between Alice
and Bob(using particle in Bob's region I)(Alice and Bob(using
anti-particle in Bob's region I)) are given by
$\rho^{W}_{AB_{I}^{+}}$ and $\rho^{W}_{AB_{I}^{-}}$ beyond single mode approximation.
Moreover, the physical states shared between Alice and
Bob(using particle in Bob's region II)(Alice and Bob(using
anti-particle in Bob's region II)) are
$\rho^{W}_{AB_{II}^{+}}$ and $\rho^{W}_{AB_{II}^{-}}$.\\
 The geometric discord of $\rho_{AB_{\mathrm{I}}^{+}}^{W}$, $
\rho_{AB_{\mathrm{I}}^{-}}^{W}$, $ \rho_{AB_{\mathrm{II}}^{+}}^{W}$
and $ \rho_{AB_{\mathrm{II}}^{-}}^{W}$ are found in Fig. 3 and Fig.
4. We observe that all quantum states show  non-vanishing quantum
mutual information and classical information even at the infinite
acceleration but the geometric discords of the quantum states such
as $\rho_{AB_{\mathrm{I}}^{-}}^{W}$ and $
\rho_{AB_{\mathrm{II}}^{+}}^{W}$ disappear at the infinite
acceleration.

\begin{figure}[h!]
\begin{center}
\includegraphics[width=7cm]{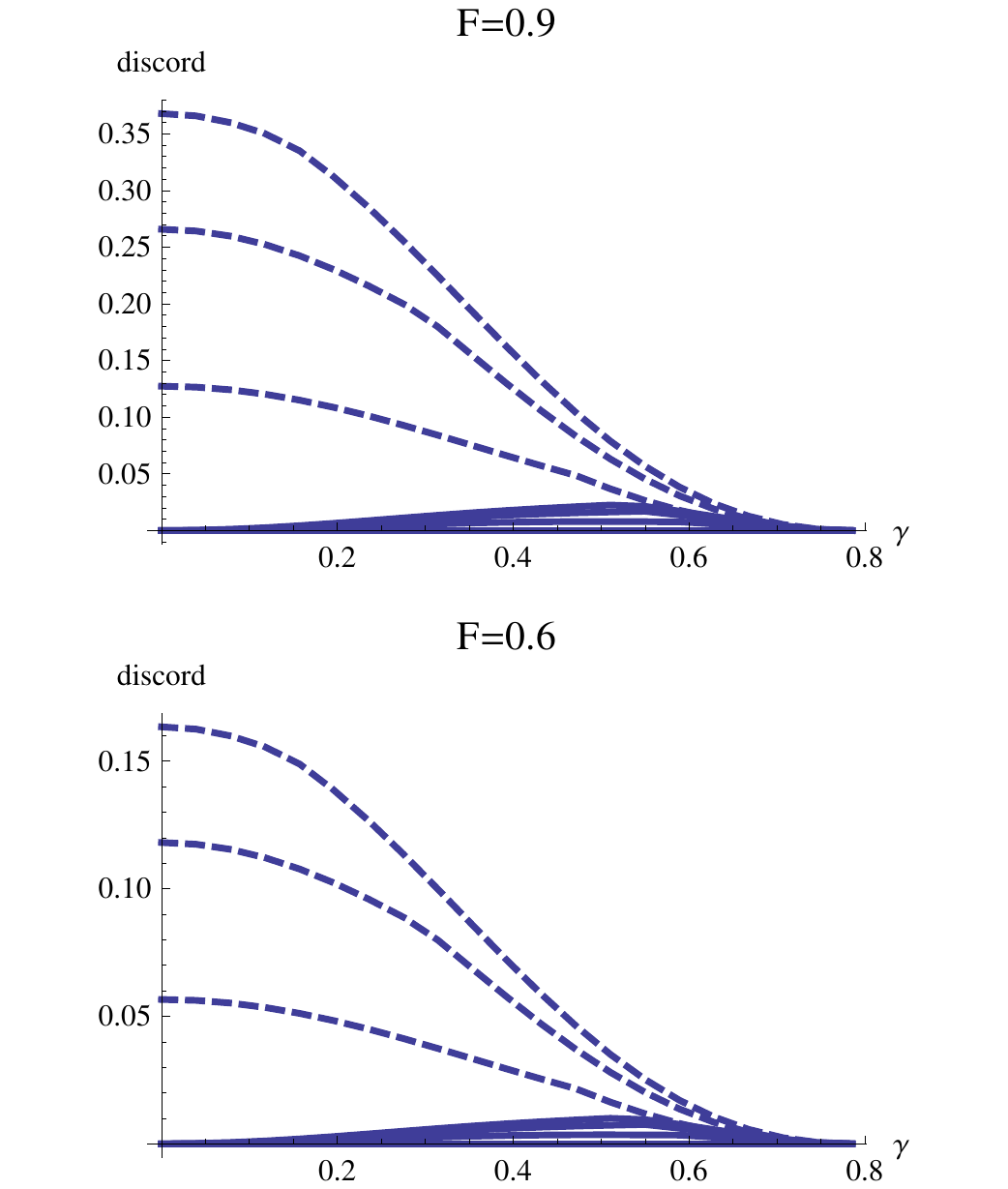}
\caption{\label{fig2}(Color online) The geometric discord of quantum
states $\rho_{AB_{\mathrm{I}}^{-}}^{W}$ and $
\rho_{AB_{\mathrm{II}}^{+}}^{W}$ when $F=0.9$ and $F=0.6$. The
solid(dotted) lines (from top to bottom) denote the geometric
discord of $\rho_{AB_{\mathrm{I}}^{-}}^{W}$ and $
\rho_{AB_{\mathrm{II}}^{+}}^{W}$ at $q_{R}=1$, $q_{R}=0.75$
$q_{R}=0.5$ and $q_{R}=0.25$ respectively. The geometric discord of
the quantum states such as  $ \rho_{AB_{\mathrm{I}}^{-}}^{W}$ and $
\rho_{AB_{\mathrm{II}}^{+}}^{W}$ disappear at the infinite
acceleration. Here $\gamma=\frac{\pi}{4}$ denotes the infinite
acceleration. }
\end{center}
\end{figure}

\section*{III. Discussion and Conclusion}
We have investigated the geometric discord of entangled states such
as $\Phi^{+}$, $\Phi^{-}$ and the Werner state for fermionic
systems. We have found that beyond the single-mode approximation,
the geometric discord for entangled quantum states of fermionic
system in accelerated frame does not vanish even at the infinite
acceleration limit if the quantum state of Alice's particle and
Bob's particle in Bob's region I or Alice's particle and Bob's
anti-particle in Bob's region II but the quantum state of Alice's
particle and Bob's anti-particle in Bob's region I or Alice's
particle and Bob's particle in Bob's region II have vanishing
geometric discord in the same infinite acceleration limit. We also
observed that the quantum discord of $\Phi^{+}$ and $\Phi^{-}$
states behaves in the analogous way.

\section*{Acknowledgement}
 This work is supported by Basic Science Research Program through the National Research
Foundation of Korea funded by the Ministry of Education, Science and
Technology (KRF2011-0027142). KLC acknowledges support from the National
Research Foundation \& Ministry of Education, Singapore.

\end{document}